\documentclass[final]{agujournal2019}
\usepackage{url} %this package should fix any errors with URLs in refs.
\usepackage{lineno}
\usepackage[finalnew]{trackchanges} %for better track changes. finalnew option will compile document with changes incorporated.
\usepackage{soul}

\journalname{JGR: Space Physics}

\begin{document}

%% ------------------------------------------------------------------------ %%
%  Title
%
% (A title should be specific, informative, and brief. Use
% abbreviations only if they are defined in the abstract. Titles that
% start with general keywords then specific terms are optimized in
% searches)
%
%% ------------------------------------------------------------------------ %%

% Example: \title{This is a test title}

\title{Evolution of a Long-Duration Coronal Mass Ejection and its Sheath Region Between Mercury and Earth on 2013 July 9-14}

%% ------------------------------------------------------------------------ %%
%
%  AUTHORS AND AFFILIATIONS
%
%% ------------------------------------------------------------------------ %%

% Authors are individuals who have significantly contributed to the
% research and preparation of the article. Group authors are allowed, if
% each author in the group is separately identified in an appendix.)

% List authors by first name or initial followed by last name and
% separated by commas. Use \affil{} to number affiliations, and
% \thanks{} for author notes.
% Additional author notes should be indicated with \thanks{} (for
% example, for current addresses).

% Example: \authors{A. B. Author\affil{1}\thanks{Current address, Antartica}, B. C. Author\affil{2,3}, and D. E.
% Author\affil{3,4}\thanks{Also funded by Monsanto.}}

\authors{N.\ Lugaz\affil{1}, R.\ M.~Winslow\affil{1}, C.\ J.~Farrugia\affil{1}}

\affiliation{1}{Space Science Center and Department of Physics, University of New Hampshire, Durham, NH, USA}
% \affiliation{2}{Second Affiliation}
% \affiliation{3}{Third Affiliation}
% \affiliation{4}{Fourth Affiliation}

%% Corresponding Author:
% Corresponding author mailing address and e-mail address:

% (include name and email addresses of the corresponding author.  More
% than one corresponding author is allowed in this LaTeX file and for
% publication; but only one corresponding author is allowed in our
% editorial system.)

% Example: \correspondingauthor{First and Last Name}{email@address.edu}

\correspondingauthor{No\'e Lugaz}{noe.lugaz@unh.edu}

%% Keypoints, final entry on title page.

%  List up to three key points (at least one is required)
%  Key Points summarize the main points and conclusions of the article
%  Each must be 100 characters or less with no special characters or punctuation and must be complete sentences

% Example:
% \begin{keypoints}
% \item	List up to three key points (at least one is required)
% \item	Key Points summarize the main points and conclusions of the article
% \item	Each must be 100 characters or less with no special characters or punctuation and must be complete sentences
% \end{keypoints}

\begin{keypoints}
\item Evolution and kinematics of a long-duration coronal mass ejection between Mercury and Earth are investigated.
\item Global and local measures of CME expansion are found to be in agreement, the CME is already wide at Mercury and expansion at Earth is typical.
\item The CME sheath at Earth is composed of a planar structure already present at Mercury and new accumulated material.
\end{keypoints}

%% ------------------------------------------------------------------------ %%
%
%  ABSTRACT and PLAIN LANGUAGE SUMMARY
%
% A good Abstract will begin with a short description of the problem
% being addressed, briefly describe the new data or analyses, then
% briefly states the main conclusion(s) and how they are supported and
% uncertainties.

% The Plain Language Summary should be written for a broad audience,
% including journalists and the science-interested public, that will not have 
% a background in your field.
%
% A Plain Language Summary is required in GRL, JGR: Planets, JGR: Biogeosciences,
% JGR: Oceans, G-Cubed, Reviews of Geophysics, and JAMES.
% see http://sharingscience.agu.org/creating-plain-language-summary/)
%
%% ------------------------------------------------------------------------ %%

%% \begin{abstract} starts the second page

\begin{abstract}
Using in situ measurements and remote-sensing observations, we study a coronal mass ejection (CME) that left the Sun on 9 July 2013 and impacted both Mercury and Earth while the planets were in radial alignment (within $3^\circ$). The CME had an initial speed as measured by coronagraphs of 580~$\pm$~20~km\,s$^{-1}$, an inferred speed at Mercury of 580~$\pm$~30~km\,s$^{-1}$ and a measured maximum speed at Earth of 530~km\,s$^{-1}$, indicating that it did not decelerate substantially in the inner heliosphere. The magnetic field measurements made by MESSENGER and {\it Wind} reveal a very similar magnetic ejecta at both planets. We consider the CME expansion as measured by the ejecta duration and the decrease of the magnetic field strength between Mercury and Earth and the velocity profile measured {\it in situ} by {\it Wind}. The long-duration magnetic ejecta (20 and 42 hours at Mercury and Earth, respectively) is found to be associated with a relatively slowly expanding ejecta at 1~AU, revealing that the large size of the ejecta is due to the CME itself or its expansion in the corona or innermost heliosphere, and not due to a rapid expansion between Mercury at 0.45~AU and Earth at 1~AU. We also find evidence that the CME sheath is composed of compressed material accumulated before the shock formed, as well as more recently shocked material.
\end{abstract}

%% ------------------------------------------------------------------------ %%
%
%  TEXT
%
%% ------------------------------------------------------------------------ %%

\section{Introduction} \label{intro}

Coronal mass ejections (CMEs) are one of the major drivers of space weather at all terrestrial planets in our solar system \cite{Futaana:2008,Edberg:2011, Jakosky:2015, Winslow:2017,Lee:2017}. They occur at the Sun on average several times per day \cite<with a frequency varying from 0.5 per day in solar minimum to 4 per day in solar maximum, see>{StCyr:2000,Yashiro:2004} and consist of a region of magnetically dominated plasma, the magnetic ejecta (ME), preceded in the majority of cases, by a dense sheath.  About half the events near 1 AU have a fast-forward shock that precedes the sheath and ejecta \cite{Schwenn:2005,Jian:2006,Jian:2018, Richardson:2010,Kilpua:2015}. However, due to the difference in orbital distance, the internal properties of CMEs are widely different at Mercury (0.31-0.47~AU) as compared to at Earth (1~AU) or Mars (1.48-1.67~AU) \cite{Bothmer:1998,Liu:2005,Jian:2008,Winslow:2015}. CMEs are known to impact planetary bow shocks and magnetopauses, at Mercury and at Earth \cite{Slavin:1981,Farris:1994,Shue:1998,Winslow:2017}, due to the presence of regions of high dynamic pressure and reduced magnetosonic Mach numbers.

The evolution of CMEs in the inner heliosphere (and, therefore, between the orbits of Mercury and Earth) is dominated by: 1) the ME expansion into the solar wind, which controls the ME magnetic field strength \cite{Bothmer:1998,Liu:2005,Gulisano:2010,Winslow:2015, Good:2018} and 2) the interaction of the CME with the solar wind, which controls the CME velocity and is often modeled as a hydrodynamic-like drag term \cite{Vrsnak:2001,Vrsnak:2010,Gopalswamy:2001b,Cargill:2004}. The physical mechanism and consequences of CME radial evolution have been primarily obtained from statistical studies \cite{Liu:2005,Winslow:2015,Janvier:2019} or numerical simulations \cite<e.g., see review by>{Manchester:2017}. The statistical studies often assume that the ME evolution can be modeled as quasi-self-similar, {\it i.\,e}.  with little changes in the ME internal properties beyond expansion and deceleration. Although some case studies indeed show a self-similar evolution \cite{Good:2015,Good:2019}, others reveal a much more complex evolution, with rotation \cite{Nieves:2012}, reconnection occurring inside the ME \cite{Steed:2011,Winslow:2016,Jian:2018} and CME-CME interaction occurring en route to Earth \cite{Lugaz:2017,Wang:2018}. Numerical simulations may be limited as well, since they typically assume that the CME can be modeled using magneto-hydrodynamics (MHD), which is appropriate for the ME but typically does not reproduce the sheath region well \cite{Lugaz:2007}. Additionally, one of the limitations of simulations is that the final result, in the form of simulated {\it in situ} measurements at Earth for example, strongly depends on the initial configuration and model used for the CME initiation as well as the solar wind model, but there has not been systematic investigations of the influence of these on the CME expansion or evolution.

Past conjunctions between the MErcury Surface, Space ENvironment, GEochemistry, and Ranging \cite<MESSENGER, see>{Solomon:2007} spacecraft at Mercury and spacecraft at 1 AU have made possible a number of detailed studies of CME evolution in the inner heliosphere  \cite{Winslow:2016,Winslow:2018,Good:2015,Good:2018,Wang:2018, Vrsnak:2019}. However, a number of physical questions regarding the evolution of CMEs have not been fully addressed by these past studies. Here, we focus on two particular questions: (1) How does the CME sheath develop?, and (2) Are long-duration MEs caused by ``nature'' (starting from the eruption at the Sun) or ``nurture'' (due to interaction or evolution)? The sheath region of fast shock-driving CMEs is often thought of being composed of the shocked material. However, depending on the radial distance where the shock forms, a portion of the sheath may be in fact associated with compressed (but not shocked) material from the early phase of the CME propagation \cite<e.g., see>{Farrugia:2008}. \citeA{deForest:2013} studied the evolution of a slow CME without a shock using remote observations from the Solar-TErrestrial RElations Observatory \cite<STEREO, see>{Kaiser:2008}  and {\it in situ} measurements and argued that the sheath is composed of coronal as well as compressed heliospheric material. Investigations of CME sheath regions using {\it in situ} data have focused on planar magnetic structures \cite{Palmerio:2016},  turbulence in the sheath \cite{Kilpua:2015}, draped field around the ME \cite{McComas:1989,Leitner:2005}, or the relation between the CME properties and that of the sheath \cite{Owens:2005}. The evolution of CME sheath with distance has also been studied through numerical simulations \cite{Savani:2012b}, revealing a complex dependence on distance and the ME radius of curvature. 

A typical ME at 1~AU lasts about 1 day and has a radial size of 0.21~AU \cite{Lepping:1995,Richardson:2010}. Long-duration magnetic ejecta at 1~AU have been studied by \citeA{Marubashi:2007}, who focused on 17 magnetic clouds measured between 1995 and 2004 lasting more than 30 hours each. They interpreted some measurements as being associated with a flank crossing of a curved flux rope (torus).  \citeA{Dasso:2007} and \citeA{Lugaz:2014} interpreted some of these events as the results of the merging of two successive CMEs with different orientations. Of note is that several of the events studied by \citeA{Marubashi:2007} have relatively large expansion at 1 AU, which brings the question of the large size or large expansion being the initial cause of the unusual measurements. CME expansion has been investigated using theoretical and statistical studies by \citeA{Bothmer:1998}, \citeA{Liu:2005}, \citeA{Demoulin:2009b}, \citeA{Gulisano:2010}, and for specific examples following the pioneer work of \citeA{Farrugia:1993}. A few studies have used remote observations \cite{Savani:2009,Nieves:2012,Lugaz:2012b} but except in a few cases \cite{deForest:2013}, the expansion in the heliosphere was not compared to that derived from {\it in situ} measurements. Recently, \citeA{Dumbovic:2018} discussed the relation between heliospheric and {\it in situ} expansion as well as expansion in the size and magnetic field strength of the magnetic ejecta. Missing from these studies is an investigation of how CME size relates to expansion measured {\it in situ}. 

To start answering our two questions, we investigate the evolution of a long-duration ME preceded by a relatively complex sheath region as these structures propagated from Mercury to Earth in mid-July 2013. In Section~\ref{sec:remote}, we discuss then combine {\it in situ} measurements and remote-sensing observations to estimate the CME kinematics in the inner heliosphere. In Section~\ref{sec:insitu}, we discuss the overall {\it in situ} measurements of the CME by MESSENGER and {\it Wind}. In Section~\ref{sec:expansion}, we focus on the ME expansion, before turning our interest to the CME sheath and shock in Section~\ref{sec:shock}. In Section~\ref{sec:conc}, we conclude.  

\section{CME Remote Observations and Kinematics}\label{sec:remote}
The 2013 July 9-13 CME event was highlighted in \citeA{Moestl:2018}, with a focus on a new space weather modeling tool and the analysis of the remote observations. Here, we summarize the main results from the coronal and heliospheric observations which relate to the orientation, direction and velocity of the CME that impacted both Mercury and Earth in mid-July 2013. Mercury and Earth were separated by $\sim 3^\circ$ in heliolongitude at the time of this CME, providing one of the closest radial conjunction events ofMESSENGER with Earth.

\subsection{Solar and Coronal Observations}

On 2013 July 9 at 14:25 UT, the STEREO-B/COR1 coronagraph (field-of-view: 1.5-4~$R_\odot$) observed an eruption from the west limb, associated with a filament eruption \cite<see details in>{Moestl:2018}. It appears fainter in STEREO-A/COR1 with a first image at 14:46 UT. LASCO observed a halo CME starting around 15:12~UT in the C2 coronagraph (field-of-view: 1.5-6~$R_\odot$) and around 16:30~UT in  the C3 coronagraph (field-of-view: 3.7-30~$R_\odot$). The CME can also be seen in the COR2-A and COR2-B coronagraphs (fields-of-view: 2.5-15~$R_\odot$) starting at 15:24 and 15:36~UT, respectively. At that time, the two STEREO spacecraft were about 140$^\circ$ away from the Sun-Earth line with STEREO-B east of the Sun-Earth line and STEREO-A west of it. The halo view from LASCO combined with the west-limb view from STEREO-B and east-limb view from STEREO-A indicates that this is an Earth-directed CME. Note that since the two STEREO spacecraft were more than 45$^\circ$ behind the solar limb, they observe this CME as a limb event. This CME is associated with a clear filament eruption as observed by SDO/AIA 304 \AA~starting around 14:00~UT. The filament was slightly east of disk center, and the eruption appears, from the SDO images to be confined. However, as is clear from the coronagraph images, a CME is associated with this filament eruption. A faint EUV wave is observed by SDO associated with this eruption, also confirming its eruptive nature.

\citeA{Hess:2017} performed a coronal reconstruction of this CME using the Graduated Cylindrical Shell (GCS) model of \citeA{Thernisien:2009} and found a coronal 3-D velocity of 526~km\,s$^{-1}$. This can be compared with the CME speed provided from published catalogs. To do so, we use the speed listed on the Coordinated Data Analysis Workshop (CDAW) SOHO/LASCO catalog \cite{Yashiro:2004}, as well as that listed on the automatic Solar Eruptive Event Detection System (SEEDS) COR-B and COR-A catalogs \cite{Olmedo:2008}. The CME is listed with a speed of 449, 410 and 507 km\,s$^{-1}$, respectively. They also list a source region of N15E05. The Community Coordinated Modeling Center (CCMC) of NASA Goddard Space Flight Center (GSFC) maintains a real-time Space Weather Database Of Notification, Knowledge, Information (DONKI) used to forecast CME hit/miss and arrival using real-time observations. NASA/CCMC/DONKI lists a CME with a speed of 600~km\,s$^{-1}$ at 21.5~$R_\odot$ with a central position of N02E10. Performing their own GCS reconstruction, \citeA{Moestl:2018} found an average coronal speed of 575 $\pm$ 60~km\,s$^{-1}$, a central position of S01W01 and a tilt of $-18^\circ$ $\pm$ 6$^\circ$. As discussed below, a CME speed close to 600~km\,s$^{-1}$ in the upper corona is likely based on transit time considerations.

\subsection{Heliospheric Observations}
Although the CME propagated more than 120$^\circ$ from the Sun-STEREO-B line and more than 140$^\circ$ from the Sun-STEREO-A line, it can be clearly seen in HI1 and, for STEREO-B in HI2 as well. As seen in time-elongation maps at a fixed position angle \cite<J-maps, see>{Davies:2009}, STEREO-B has a view up to about 35$^\circ$ elongation, whereas STEREO-A does not observe the CME in HI2 field-of-view, and barely past 15$^\circ$. While it may seem surprising that a CME can be observed into HI2 field-of-view so far from the Sun-STEREO line, this possibility was raised relatively early in the mission \cite{Lugaz:2012a}. This is due to a combination of the wide nature of the CME density structure and the relatively low increase in scattering efficiency away from the Thomson sphere \cite{THoward:2012c}.  However, at large observing angles, the convolution between CME shape, kinematics and the observing geometry becomes inextricable \cite{Lugaz:2013a,Moestl:2018,THoward:2013b,Liu:2016} and methods to derive the CME properties have not been carefully tested for these conditions. We use measurements made from the STEREO-B J-maps to determine the CME kinematics and compare with the results from \citeA{Moestl:2018}.

 %%%%%%%%%%%%%%%%%%%%%%%
\begin{figure}[tb]
\centering
{\includegraphics*[width=.98\linewidth]{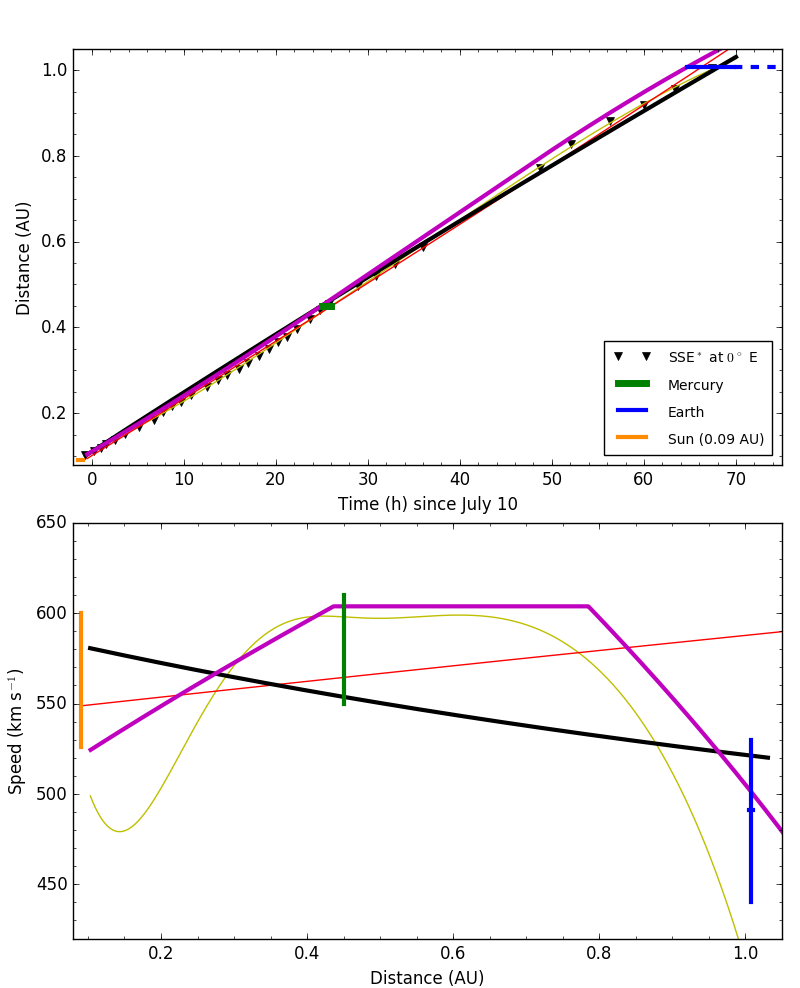}}
\caption{Position {\it vs}.\ time (top) and speed {\it vs}.\ distance (bottom) of the CME front as derived from HI measurements. Inverted triangles: Radial position derived from elongation measurements using the SSE model. The lines show the fit to the position using different assumptions. The horizontal (top) and vertical (bottom) lines indicate the arrival time at Mercury (green) and at Earth (blue) as well as the initial speed estimated from coronagraphs (orange). The different line corresponds to a linear function (red), a step-wise linear function (magenta), a spline fit (yellow) and the fit with a drag-based model (black).The heavy lines (horizontal in the top panel, vertical in the bottom panel) show the uncertainty for the CME initial leading edge speed and position (yellow), the shock arrival time (top green) and CME leading edge speed at MESSENGER (bottom green) and the sheath arrival time and speed at Earth (blue).}
\label{fig:HI}
\end{figure}
%%%%%%%%%%%%%%%%%%%%%%%% 

\subsection{Combining Remote and In Situ Observations to Determine CME Kinematics}\label{sec:truc}
Various methods have been developed after the launch of the STEREO mission to derive the CME kinematics and direction of propagation from HI observations, and, hence, determine their arrival times at various locations. For remote measurements from one spacecraft only, the main methods assume a constant CME width, constant propagation speed and constant direction of propagation; these are: the Fixed-$\Phi$ \cite{Rouillard:2008} that assumes a negligible CME width, the Harmonic Mean \cite<HM, see>{Lugaz:2009c} that assumes a 90$^\circ$ half-width and the Self-Similar Expansion \cite<SSE, see>{Davies:2012} for which the half-width can be fixed to any value. In general, the SSE method has proven the most reliable, but it is important to compare the results of these various methods to understand the uncertainties in deriving kinematics and arrival time associated with the unknown width of the CME. Because STEREO-A did not observe the CME into the HI-2 field-of-view, we rely on these single-spacecraft techniques rather than more accurate stereoscopic techniques. Here, we use these techniques in the ecliptic plane and determine the radial speed of the CME front and its direction given as a longitude with respect to the Sun-Earth line.

Through the Heliospheric Cataloguing, Analysis and Techniques Service (HELCATS) project \cite{Barnes:2019}, the CME average speed and propagation direction was determined for the Fixed-$\Phi$ and SEE techniques. 
Using the Fixed-$\Phi$ method or the SSE method with a half-width of 30$^\circ$, the CME speed is found to be 431 $\pm$ 12~km\,s$^{-1}$ and 513 $\pm$ 21~km\,s$^{-1}$, respectively. The CME direction is found to be $-25~\pm~1^\circ$ and $-2~\pm~2^\circ$, in relative agreement with the direction obtained in the corona between  $-12^\circ$ and $1^\circ$.
 
Using the same HI observations from STEREO-A, it is possible to constrain the speed and direction by making use of the known arrival times at MESSENGER (at 0.450~AU) with the sheath passing between 01:05 and 04:05~UT on July 11 and at {\it Wind} (at 1.008~AU) with the dense sheath passage between 16:43 on July 12 and 05:30 on July 13. Using the HM method, the arrival time at MESSENGER and at {\it Wind} is always significantly too late compared to the measured ones for any CME directions between $-20^\circ$ and 0$^\circ$ {\it i.\ e}.\ for any CME direction consistent with the coronagraphic measurements. As such, the HM method does not provide reliable results for this CME.

Using the SSE method, the half-width of the CME can be adjusted to match the measurements. However, as pointed out in \citeA{Davies:2013}, the effect of the half-width and CME direction cannot be disentangled. We consider half-widths between 25$^\circ$ and 45$^\circ$ to be realistic and direction between $-20^\circ$ and 0$^\circ$. A good-fit to the arrival times at MESSENGER and {\it Wind} can be obtained for a half-width of 28$^\circ$ and a direction of $0^\circ$. The match to both MESSENGER and {\it Wind} arrival is significantly better using the SSE method with a half-width of 28$^\circ$ than using the Fixed-$\Phi$ method. 

The sheath at {\it Wind} is measured with a speed varying between 440 and 540~km\,s$^{-1}$. As discussed earlier, the initial CME speed at 0.1~AU is estimated to be between 526 and 600~km\,s$^{-1}$. From this, it is already evident that the CME did not decelerate much in the interplanetary (IP) space. In fact, the average transit speed from first remote detection by STEREO-B/COR1 to Mercury is 534~km\,s$^{-1}$, from first detection to Earth is 556~km\,s$^{-1}$ and from Mercury to Earth, it is 570~km\,s$^{-1}$.
  
Using the SSE method for a width of 28$^\circ$ and a direction of propagation along the Sun-Earth line, the CME time-elongation plot is well-fitted by a CME with a nearly constant speed, i.\ e.\ an initial speed of 550~km\,s$^{-1}$ and speed at 1~AU of 585~km\,s$^{-1}$ as shown with the red line in Figure~\ref{fig:HI} (bottom panel). More complex spline fitting (see yellow line in Figure~\ref{fig:HI}) does not return more realistic variations of the CME speed, but typically one with a maximum speed of 585-645~km\,s$^{-1}$ reached between 0.4 and 0.7~AU. These fits indicate that the CME may have accelerated up to 0.4 to 0.5~AU and decelerated later on. The long-lasting acceleration would be a relatively unexpected behavior, as most CMEs reach their peak speed in the corona.

We also attempted to fit the initial speed, arrival times and arrival speeds using the drag-based method following \citeA{Vrsnak:2013} and \citeA{Hess:2017} as shown with a black line in Figure~\ref{fig:HI}. We use an asymptotic solar wind speed of 375~km\,s$^{-1}$, in agreement with the speed measured upstream of the CME-driven shock at 1~AU.  For the CME to impact MESSENGER at the measured time, the initial CME speed at 0.1~AU must be 570 $\pm~10$~km\,s$^{-1}$ and the drag parameter is extremely low at 6 $\pm$ 3 $\times 10^{-9}$~km$^{-1}$ or lower than the lowest recommended value. For example, \citeA{Vrsnak:2013} consider that the drag parameter should be between 10$^{-8}$ and 10$^{-5}$~km$^{-1}$. Using the larger initial speed and the larger drag parameter, the derived speed at MESSENGER is 554~km\,s$^{-1}$ and the \add{predicted} speed at 1~AU is 521~km\,s$^{-1}$. \change{with an arrival time}{The predicted arrival time of the CME front occurs} during the sheath passage but about 3 hours after the shock arrival (black line in Figure~\ref{fig:HI}). Under the assumption of a series of constant acceleration periods \cite<see examples in>{Wood:2009c} and following the behavior found using the spline fitting, a good match can be obtained with a step-wise linear function using three periods (see magenta line in Figure~\ref{fig:HI}): constant acceleration of 0.9~m\,s$^{-2}$ for the first 24 hours (starting at a distance of 0.11~AU, i.\ e.\ 23.5~$R_\odot$), constant speed for the next 24 hours and constant deceleration of $-1.7$~m\,s$^{-2}$ until 1~AU. This assumes that the initial CME speed at 0.11~AU is 526~km\,s$^{-1}$. The maximum CME speed (also its speed at MESSENGER) is 604~km\,s$^{-1}$ and the speed at 1~AU is 501~km\,s$^{-1}$, close to the average speed in the sheath (491~km\,s$^{-1}$). 

As a summary, based on remote observations, the results of fitting methods, and arrival time constraints, the CME front speed was likely to be of the order of 550--610~km\,s$^{-1}$ ({\it i.\ e}. 580 $\pm$ 30~km\,s$^{-1}$) at MESSENGER, indicating a possible small acceleration between the upper corona and Mercury's orbit or at the very least a very small deceleration. The speed measured {\it in-situ} at 1~AU (average sheath speed of 491~km\,s$^{-1}$) and the average transit speed of 556~km\,s$^{-1}$ confirm that the CME did have very little deceleration in the heliosphere. This can also be seen by the fact that the CME transit speed from first detection in COR1 to MESSENGER is lower by about 6\% than the transit speed from MESSENGER to 1~AU. Figure~\ref{fig:HI} shows the HI measurements and their fits.

\section{In Situ Interplanetary Measurements}\label{sec:insitu}

\subsection{{\it Wind} Measurements at L1}
\begin{figure}[tb]
\centering
{\includegraphics*[width=.98\linewidth]{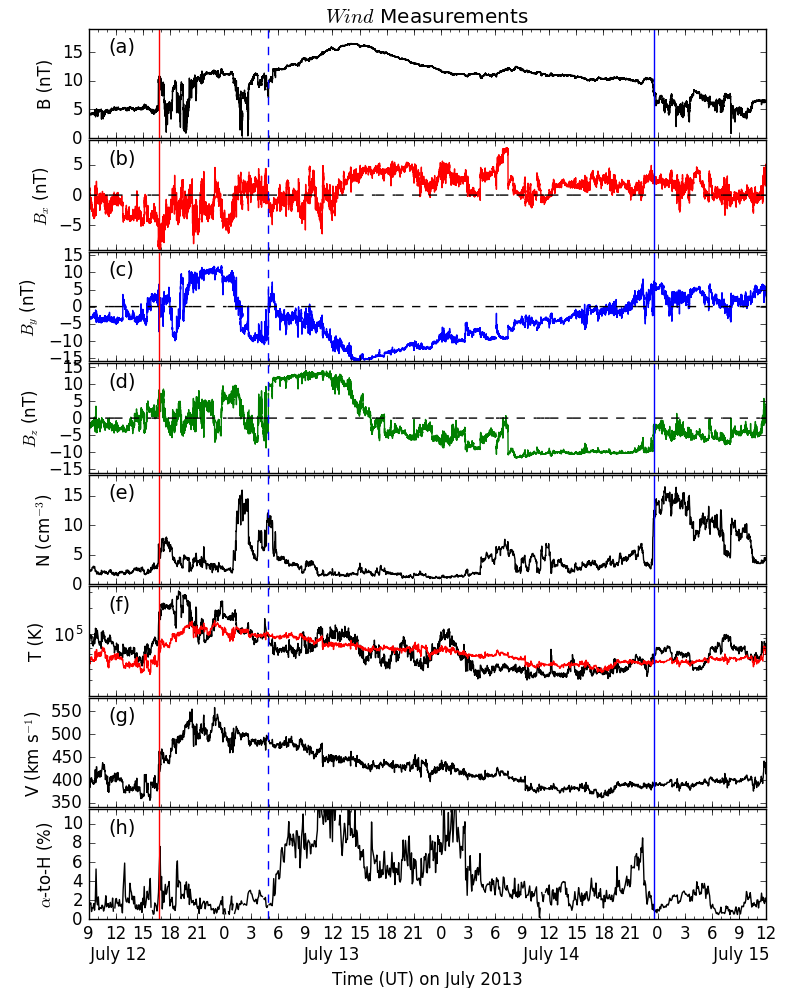}}
%{\includegraphics*[width=.48\linewidth]{Messenger.png}}
\caption{CME Measurements at {\it Wind}. The panels show from top to bottom, the magnetic field strength (a), the $x$, $y$ and $z$ components in GSE (b-d), the proton density (e), temperature (red, expected temperature from \citeA{Lopez:1987}) (f), the velocity (g), and the $\alpha$-to-proton number density ratio (h). The red, dashed blue and \add{solid} blue vertical lines indicate the shock arrival, the beginning and end of the magnetic ejecta.}
\label{fig:Wind_full}
\end{figure}

Although the CME first impacted MESSENGER before impacting {\it Wind}, we describe the {\it in situ} measurements made at L1 by {\it Wind} first, as the availability of plasma measurements make\add{s} it easier to identify the different start and end times of the CME sub-structures.
At 16:43 UT on July 12, a fast forward shock impacted {\it Wind} (see red vertical line in Figure~\ref{fig:Wind_full}). The \url{IPshocks.fi} database of \citeA{Kilpua:2015} lists a quasi-perpendicular ($\theta_{Bn}$ = 79$^\circ$) shock with a speed of 493~km\,s$^{-1}$, a density and magnetic field compression ratios of 1.6 and 2.0, respectively, an upstream solar wind and fast magnetosonic speeds of 377~km\,s$^{-1}$ and 85~km\,s$^{-1}$, respectively. This results in a fast magnetosonic Mach number of 2.6, relatively typical for IP shocks at 1~AU. Figure~\ref{fig:Wind_full} shows the measurements by {\it Wind} including the expected temperature from \citeA{Lopez:1987} in red in  panel (f) and the $\alpha$-to-proton number density ratio in the last panel. 

The dense sheath lasts until 4:55 UT on July 13 with the beginning of the ME characterized by the start of a $B_z$ positive period (panel (d) in Figure~\ref{fig:Wind_full}),  a drop of the measured temperature (panel (f) in Figure~\ref{fig:Wind_full}) below the expected temperature of \citeA{Lopez:1987} and the beginning of the bi-directional suprathermal electrons (not shown). A slightly later start time of the ME at 5:45 UT could be chosen to match the drop in density (panel (e) in Figure~\ref{fig:Wind_full}), but as is explained later we chose the 4:55 UT time as more appropriate to compare with MESSENGER data. The $\alpha$-to-proton number density ratio (panel (h) in Figure~\ref{fig:Wind_full}) increases from $\sim$2\% to $\sim$6\% between 4:45 UT and 6:15 UT, which also confirms the start of the ME occurs during this time.

The end of the ME detection is relatively clear with a discontinuity-like feature in the proton density and magnetic field at 23:31 UT on July 14. This is marked by the solid blue line in Figure~\ref{fig:Wind_full}, where the discontinuity is particularly visible in the total magnetic field (panel a), $B_z$ component of the magnetic field (panel d) and density (panel e). The typical duration of a ME at 1~AU is 18--26 hours \cite{Gopalswamy:2015,Nieves:2018}, while the sheath duration is typically one third of that of the ejecta or about 6--9 hours \cite{Jian:2018}. The sheath duration of 12.2 hours is somewhat large for CMEs measured at 1~AU, while the ME duration of 42.6 hours is significantly longer than average.

 %%%%%%%%%%%%%%%%%%%%%%%
%%%%%%%%%%%%%%%%%%%%%%%% 

The CME is listed in the database of \citeA{Richardson:2010} and that of \citeA{Nieves:2018}, and the ME is identified as a magnetic cloud by \citeA{Lepping:2018}. The long-duration, relatively complex rotation, and period of enhanced density about 22 hours after the start of the ME are somewhat reminiscent of the complex ejecta described in \citeA{Lugaz:2014}. \citeA{Nieves:2018} categorize this event as ``complex''. The magnetic cloud radial size (diameter), as reported in \citeA{Lepping:2018} is 0.475~AU (more than twice that of a typical magnetic cloud). A minimum variance analysis of the magnetic field data in geocentric solar ecliptic (GSE) coordinates data \cite{Sonnerup:1967,Sonnerup:1998} returns a low-inclined magnetic ejecta with an axis in the $y$ direction (longitude $\sim 102^\circ$ and latitude of $-2^\circ$), consistent with the GCS fitting of the remote-sensing data with a lower tilt as performed by \citeA{Hess:2017}. 
%The dimensionless impact parameter is $-0.13$, indicating a crossing close to the axis and the cone angle (the complementary angle to the position angle of \citeA{Janvier:2013}) is 84$^\circ$, indicating a crossing close to the nose. The quality of the fit is reported as ``poor''. 

Note that the end time of the ME in \citeA{Lepping:2018} is pushed to 7:20 UT on July 15, which we consider unlikely in light of the discontinuity occurring at 23:31 on July 14 (marked by the solid blue line). On the other hand, \citeA{Nieves:2018} report a start time of the ME at 20:24 UT on July 12, which corresponds to the peak velocity and the beginning of 5-hour period of smooth and elevated magnetic field. The lack of bi-directional suprathermal electrons and low value of the $\alpha$-to-proton number density ratio make it unlikely that this is the actual start of the ME. Further discussion about the sheath in the next section highlights that this feature is likely to be a planar structure within the sheath.

\subsection{MESSENGER Measurements at Mercury}\label{sec:Merc}
\citeA{Winslow:2015} presented a database of CMEs measured at Mercury by MESSENGER, and we follow their approach to determine the CME arrival time, as well as the procedure of \citeA{Winslow:2013} to identify the crossings of Mercury's bow shock and magnetopause. MESSENGER provides magnetic field measurements  but no reliable plasma measurements in the solar wind. 

During this time period, 1) MESSENGER's orbital period was 8 hours with an approximately dawn-to-dusk orbit, 2) Mercury's bow shock is most easily identified during the outbound part of the orbit, and, 3) for some orbits there are only 30 minutes of pristine interplanetary magnetic field (IMF) measurements as MESSENGER remains for extended periods in the planetary magnetosheath. In the orbit just before the arrival of the CME, there are however $\sim$ 3.5 hours of IMF measurements, and during the following 4 orbits, periods of $\sim$ 4.5, 0.5, 5.25 and 3.75 hours when MESSENGER is in the solar wind. As MESSENGER remains for extended period in Mercury's magnetosheath, where the IMF is modified by Mercury's bow shock, we extrapolate the IMF measurements by dividing the measured magnetic field inside the magnetosheath by the jump experienced through the bow shock for the magnitude and each individual components. This procedure can be performed when MESSENGER is in the magnetosheath 1) between an inbound and outbound bow shock crossings without any magnetopause crossing,  2) between the bow shock crossing and $\sim 10-20$ minutes before the subsequent magnetopause crossing, or 3) between $\sim 30$ minutes after a magnetopause crossing and the subsequent bow shock crossing. This allows us to retrieve an additional $\sim$ 4 hours of IMF measurements, especially for the second orbit which was heavily perturbed by the CME passage and for which there are almost no pristine IMF measurements but numerous periods in-between bow shock crossings. A somewhat similar approach was taken in \citeA{Wang:2018} to recover pristine CME measurements at 1 AU for an event through which an overtaking shock was propagating.

\begin{figure}[tb]
\centering
{\includegraphics*[width=.98\linewidth]{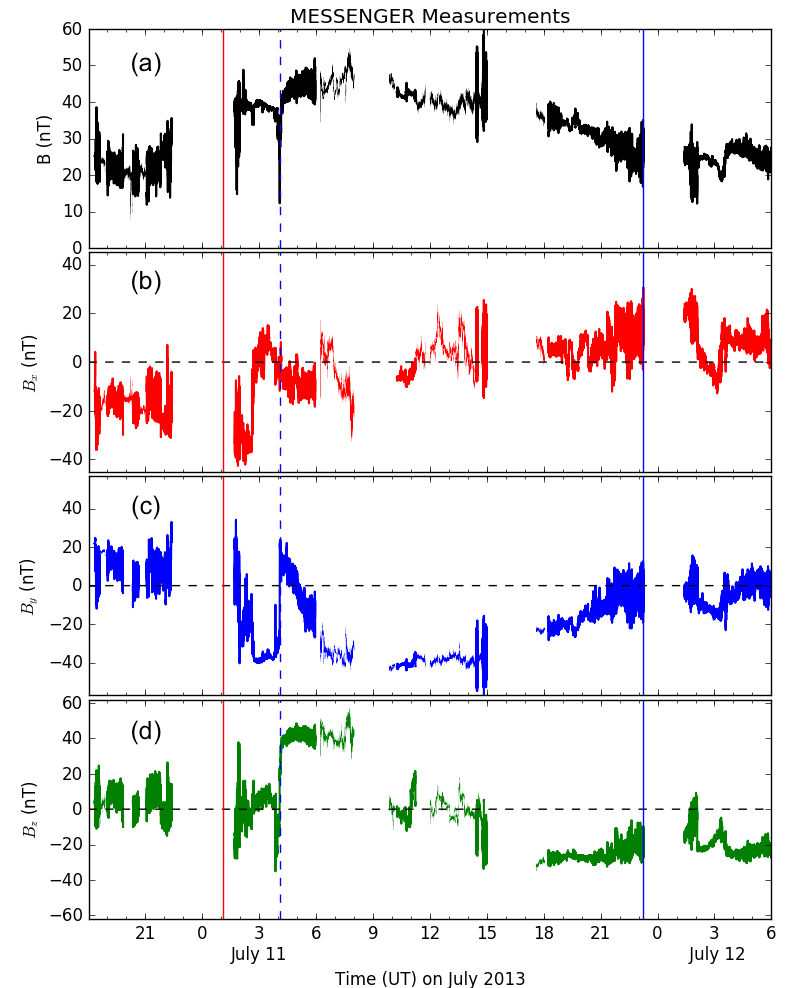}}
\caption{CME Measurements at MESSENGER. The panels show from top to bottom, the magnetic field strength and the $x$, $y$ and $z$ components in pseudo-GSE coordinates. The red, dashed blue and \add{solid} blue vertical lines indicate the shock arrival, the beginning and end of the magnetic ejecta. The thinner lines indicate data recovered from inside Mercury's magnetosheath, whereas the thick lines indicate data taken while MESSENGER was directly measuring the IMF.}
\label{fig:Messenger}
\end{figure}

In Figure~\ref{fig:Messenger}, the  data recovered from Mercury's magnetosphere is plotted with a thin line, while the data measured in the solar wind is plotted with a thick line. The gaps in the plotted data correspond to times  1) when MESSENGER was in Mercury's magnetosphere, i.e. between an inbound and an outbound magnetopause crossings, or 2) when MESSENGER was in Mercury's magnetosheath but the procedure to recover the IMF could not be applied following the conditions detailed above, i.e. typically in close proximity to a magnetopause crossing. When there is no clear jump in one of the magnetic field components through the bow shock, we do not recover the IMF measurements for that particular component, resulting in different data gaps for the various magnetic field components. Data in the heliosphere away from the Sun-Earth line are typically plotted in RTN coordinates. Here, we plot it in pseudo-GSE, which corresponds to $-R$, $-T$, $N$ to make an easy comparison with {\it Wind} data.

\begin{figure}[tb]
\centering
{\includegraphics*[width=.98\linewidth]{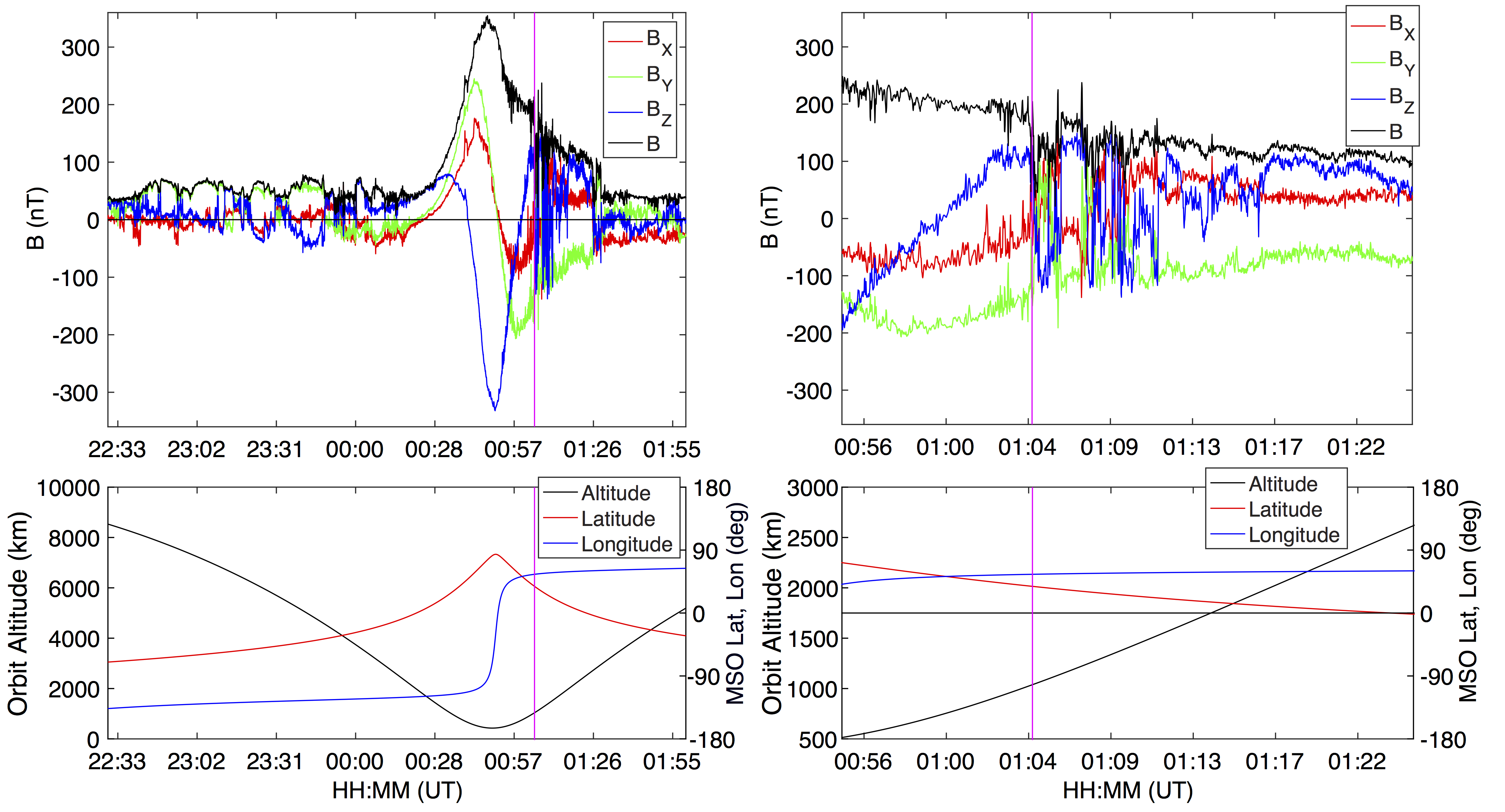}}
\caption{MESSENGER measurements of Mercury's magnetosphere during the arrival of the CME-driven shock from 22:30 UT July 10 to 02:00 UT July 11. The top panels show the magnetic field measurements in Mercury Solar Orbital (MSO) coordinates, with $X$ sunward, $Z$ perpendicular to the orbital plane and  $Y$ completing the right-handed, orthonormal system. The bottom panels show the spacecraft altitude, MSO latitude and longitude. The left-hand side shows the entire magnetospheric crossing during the particular orbit, while the right-hand side shows a zoom-in around the time of the shock arrival. In all panels, the shock arrival is marked with the magenta line. The first outbound bow shock crossings occurs around 01:26 UT on July 11.}
\label{fig:shock_arrival}
\end{figure}

Although MESSENGER was inside the magnetosphere when the CME sheath and shock arrived, the effect of the shock can clearly be seen in the compression of the magnetosphere and this allows for the identification of the shock arrival as occurring at 01:05 UT on July 11. MESSENGER measurements of Mercury's magnetosphere during the arrival of the CME-driven shock from 22:30 UT July 10 to 02:00 UT July 11 are shown in Figure~\ref{fig:shock_arrival}. The top panels show the magnetic field measurements in Mercury Solar Orbital (MSO) coordinates, with $X$ sunward, $Z$ perpendicular to the orbital plane and  $Y$ completing the right-handed, orthonormal system. The bottom panels show the spacecraft altitude, MSO latitude and longitude. The shock arrival is marked with the magenta line. It corresponds to the magnetosphere being compressed and MESSENGER temporarily exiting the magnetosphere for the magnetosheath. Further discussion of the shock is given in Section~\ref{sec:sheath}. Shortly after MESSENGER exits Mercury's magnetosphere on that orbit, the magnetic field becomes well organized and \citeA{Winslow:2015} report a starting time of the ME of 01:57 UT on July 11. The CME end time is listed as 23:14 UT on July 11, which corresponds to MESSENGER entering into Mercury's magnetosphere three orbits ($\sim$ 24 hours) after the shock arrival. 

These magnetic field measurements plotted in Figure~\ref{fig:Messenger} can be analyzed with the knowledge gained from the near-Earth's measurements (see Figure~\ref{fig:Wind_full}) that include plasma observations, in particular to determine the beginning of the ME. The time period between 01:57 UT and 04:05 UT on July 11 is highly reminiscent of the period just preceding the beginning of the ME at Earth. We consider this region to be a likely planar magnetic structure, part of the CME sheath.  A planar magnetic structure is one where the magnetic field varies in a plane \cite{Nakagawa:1989,Farrugia:1990}, and is often measured in CME sheaths \cite{Kataoka:2015,Palmerio:2016}.
To be consistent in the identification of the sheath at Mercury and Earth, we consider that the ME starts at MESSENGER at 04:05 UT (vertical blue dashed line in Figure~\ref{fig:Messenger}), for a duration of at least 19.15 hours (with some uncertainty due to the end time, as shown with the solid vertical blue line in Figure~\ref{fig:Messenger} corresponding to MESSENGER entering Mercury's magnetosphere). Choosing this time for the end of the sheath at MESSENGER ensures that the discontinuity where $B_y$ and $B_z$ turn from negative to positive (see panels c and d of Figures~\ref{fig:Wind_full} and \ref{fig:Messenger}) corresponds to the start of the ME at both locations.

A minimum variance analysis of the ME with the start and end times at 04:05 and 23:14 UT on July 11 returns an orientation of the axis within about 5.5$^\circ$ from the $y$-direction. The ratio of the middle to the lowest eigenvalues is 5.5, indicating that a reliable estimate of the axis direction can be obtained. The CME orientation at Mercury is consistent with the minimum variance analysis results from the {\it Wind} data. A visual inspection of the ME at Mercury and Earth clearly confirms that the global orientation remains relatively unchanged between the two planets, with a North-West-South ejecta following the classification of \citeA{Bothmer:1998} at both locations (meaning that the magnetic field is northward in the front half of the ejecta, southward in the back half and the axial field is in the west direction). This can be seen in panels c and d of Figures~\ref{fig:Wind_full} and \ref{fig:Messenger}, with the positive-to-negative rotation of $B_z$ and the negative $B_y$ inside the ME at both locations. We further compare the MESSENGER and {\it Wind} measurements of the CME in the following section.

\section{CME Expansion}\label{sec:expansion}

Having CME measurements from two spacecraft in longitudinal conjunction, we are able to study the CME expansion both locally at 1 AU and globally between Mercury and Earth. \citeA{Demoulin:2009b} and \citeA{Gulisano:2010} developed a theoretical paradigm with a dimensionless expansion parameter $\zeta$ defined as: 
$$\zeta = \frac{\Delta V}{\Delta t} \frac{d}{V_c^2},$$
where $\frac{\Delta V}{\Delta t}$ is the slope of the velocity profile inside the ME, $d$ is the heliospheric distance at which the measurementsare made, and $V_c$ is thespeed of the ME center. This dimensionless parameter takes into consideration that faster and longer CMEs have higher expansion speeds. $\zeta$ is a measure of the local expansion and has been found to be around 0.8 for most isolated CMEs \cite{Demoulin:2010b}. Based on theoretical and statistical analyses, it is thought to be relatively independent of distance and CME parameters, and may therefore provide a measure of the global expansion of the CME. In particular, if $\zeta$ is constant, the CME radial size is expected to grow as $r^\zeta$ and the magnetic field to decrease as $r^{-2\zeta}$, with $r$ being the heliocentric radial distance \cite<see also discussion in>{Dumbovic:2018}. 

 %%%%%%%%%%%%%%%%%%%%%%%
%%%%%%%%%%%%%%%%%%%%%%%% 

\begin{figure}[tb]
\centering
{\includegraphics*[width=.98\linewidth]{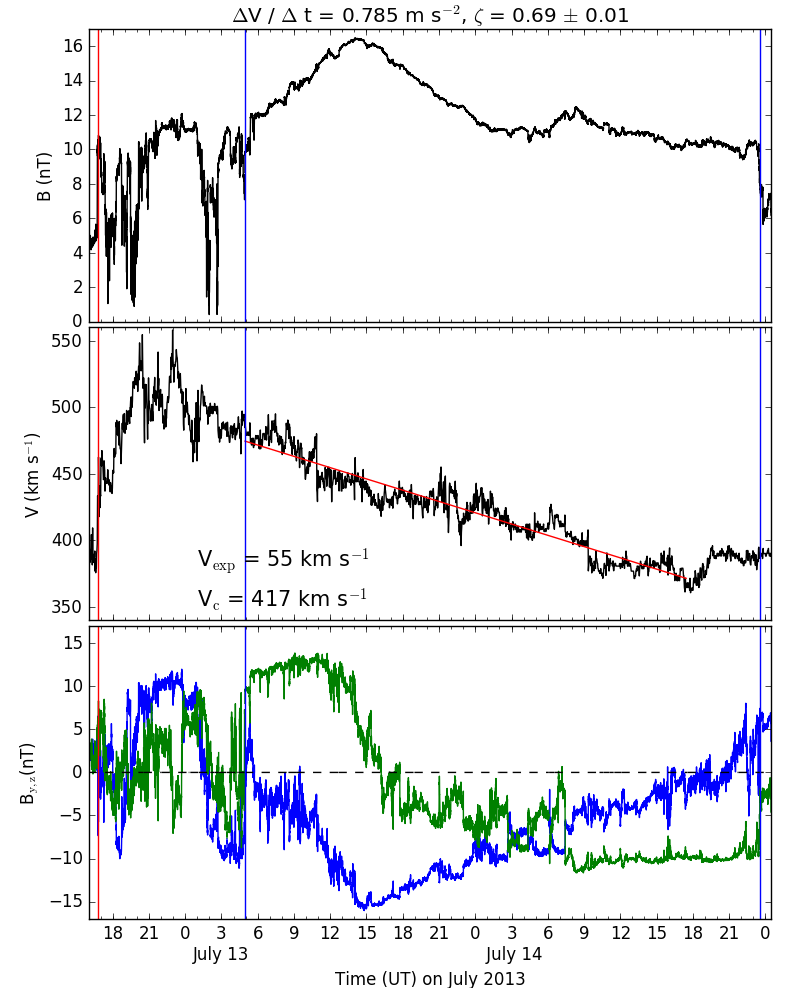}}
\caption{Fitting of the CME speed as measured by {\it Wind} with a linear function (red line in middle panel) to obtain $\zeta$. The panels show from top to bottom the magnetic field strength, proton velocity and $B_y$ (blue) and $B_z$ (green) components of the magnetic field. The vertical bars show from left to right the shock arrival, the start and end times of the ME. The speed at the center of the ME and the ME expansion speed (half the difference between the speed of the ME at the front and back boundaries) are indicated in the Figure as $V_c$ and V$_\mathrm{exp},$ respectively.}
\label{fig:expansion}
\end{figure}

Figure~\ref{fig:expansion} shows the fitting to the CME speed profile to determine the slope $\frac{\Delta V}{\Delta t}$. The slope and center speed values depend somewhat on the choice of the ME start and end boundaries. Here, we do not follow the procedure of \citeA{Gulisano:2010} of balancing the azimuthal magnetic field to determine the boundaries but use the boundaries identified in the previous section, {\it i.\ e}. a ME lasting at {\it Wind} from 4:55 UT on July 13 to 23:31 UT on July 14. We find that $\zeta = 0. 69 \pm 0.01$ depending on the exact choice of boundaries. This is relatively typical, albeit on the lower end, as \citeA{Demoulin:2010b} found that $\zeta = 0.81 \pm 0.19$ for unperturbed magnetic clouds at 1 AU. 

The magnetic field inside the ME at Earth reached a maximum value of $B_{\max,Earth} =$ 16.5~nT and an average value of $B_{av,Earth} =$ 13~nT. At Mercury, it reaches a maximum of $B_{\max,Mercury} =$ 59~nT and an average value of $B_{av,Mercury} =$ 37~nT. We define $\alpha$ as the exponent of the magnetic field decrease inside the ME between Mercury and Earth, {\it i.e}. $B_\mathrm{Earth} = B_\mathrm{Mercury}\, r_M^{\alpha}$, where $B_\mathrm{Earth}$ and $B_\mathrm{Mercury}$ are the magnetic field strength at Earth and Mercury, respectively and $r_M$ is the location of Mercury in AU (here 0.45). With this definition, we find that $\alpha_{\max} = -1.60$ for the maximum magnetic field and $\alpha_{av} = -1.31$ for the average magnetic field. This is relatively consistent with the value of $2\zeta = 1.38$ found at 1~AU for this ME and is consistent with past statistical studies \cite{Bothmer:1998,Liu:2005,Gulisano:2010,Winslow:2015}, albeit, once again, a relatively slower magnetic field decrease with distance than the typical value of $\alpha_{\max}$ of$-1.8$.

 %%%%%%%%%%%%%%%%%%%%%%%
%%%%%%%%%%%%%%%%%%%%%%%% 

 %%%%%%%%%%%%%%%%%%%%%%%
%%%%%%%%%%%%%%%%%%%%%%%% 

The radial size of the ME at Earth can be estimated by simply multiplying the average ME velocity, 417~km\,s$^{-1}$ by the ME duration, 42.5 hours, which gives 0.425 AU. This is likely a slight overestimation due to the aging effect (the CME expands as it passes over {\it Wind}). The expected size can be first estimated at Mercury using the value of $\zeta$ found. Doing so, the expected size of the CME at Mercury is 0.245 AU. Using the size at 1~AU and half of the exponent of the magnetic field expansion between Mercury and Earth, the expected size is 0.24  $\pm 0.015$~AU, a similar estimate. 

We can then relate this expected size based on the expansion and the size at 1 AU to an expected ejecta duration using the estimated CME front speed of 580 $\pm 30$~km\,s$^{-1}$ (see Section~\ref{sec:truc}) at Mercury. Simply dividing the expected ejecta duration by the estimated speed results in an estimate of the expected ME duration at MESSENGER of 17.5 $\pm 1.5$ hours. The ME duration at MESSENGER is in fact found to be 19.15 hours, on the high end of this estimate, but consistent with a lower average speed than the CME front. This indicates that the large size of the CME at 1 AU is not due to a large expansion between Mercury and Earth but to a large size already in the innermost heliosphere. If anything, the expansion appears to be typical or slightly smaller than typical. We note that the increase in the duration of the ME between Mercury and Earth is by a factor of 2.22. 

\section{CME Sheath and CME-Driven Shock}\label{sec:shock}

\subsection{CME Shock}
The sheath near Earth presents a relatively unusual velocity profile (see panel g of Figure~\ref{fig:Wind_full}). The shock is only associated with an increase of the velocity to 460~km\,s$^{-1}$, whereas the ME leading edge is at 470~km\,s$^{-1}$. In addition, a significant part of the sheath has a much larger velocity reaching up to 540~km\,s$^{-1}$.  The fast magnetosonic speed ahead of the shock is 85~km\,s$^{-1}$ and the solar wind speed is at 377~km\,s$^{-1}$. This is one case where the ME average speed of 417~km\,s$^{-1}$ is not enough to drive a shock but the ME leading edge speed, taking into consideration the expansion, is faster than the fast magnetosonic speed in the solar wind frame. Additional such examples have been discussed in \citeA{Lugaz:2017b}. A central question we want to address now is whether or not such a ME was able to drive a shock at Mercury's distance. 

We have estimated the CME leading speed to be 580 $\pm$ 30~km\,s$^{-1}$ at Mercury. Assuming that the slow solar wind speed increases as $\sqrt{\ln({r})}$ following the original derivation by~\citeA{Parker:1958}, the {\it in situ} measurements of the solar wind at Earth implies a solar wind speed at Mercury of about 340~km\,s$^{-1}$. Following the method described in \citeA{Winslow:2017}, we estimated the solar wind Mach number and the dynamic pressure using the Newtonian approximation of pressure balance at Mercury's magnetopause. For the orbit before the arrival of the CME, the Mach number is about 3.6, which corresponds to a solar wind Alfv{\'e}n speed at Mercury of $\sim$95~km\,s$^{-1}$, a bit higher than that at Earth. 

We now have estimates of the CME front speed, the solar wind speed, and the Alfv{\'e}n or fast magnetosonic speed before the CME. This gives us an estimate of the Mach number associated with the CME front speed of 2.5. As such, we find that a shock should be present at Mercury. Further discussion of the shock is given in the next section.

\subsection{CME Sheath}\label{sec:sheath}
After having clarified the existence of a CME-driven fast forward shock at Mercury, we turn our attention to the development of the sheath. \citeA{Janvier:2019} studied the average sheath duration increase from Mercury to Venus and Earth's orbits and found that the typical sheath duration increases by a factor of 5 from Mercury to Earth and the ratio of sheath to magnetic ejecta durations doubles between these two locations. Here, we find that the duration of the sheath increases by a factor 4.9 and the ratio of sheath to ejecta duration increases by a factor of 2.3, both relatively consistent with their results. This holds true even though the magnetic ejecta is more than twice longer than typical both at Mercury and Earth, whereas the sheath duration is closer to typical.

\begin{figure}[tb]
\centering
{\includegraphics*[width=.98\linewidth]{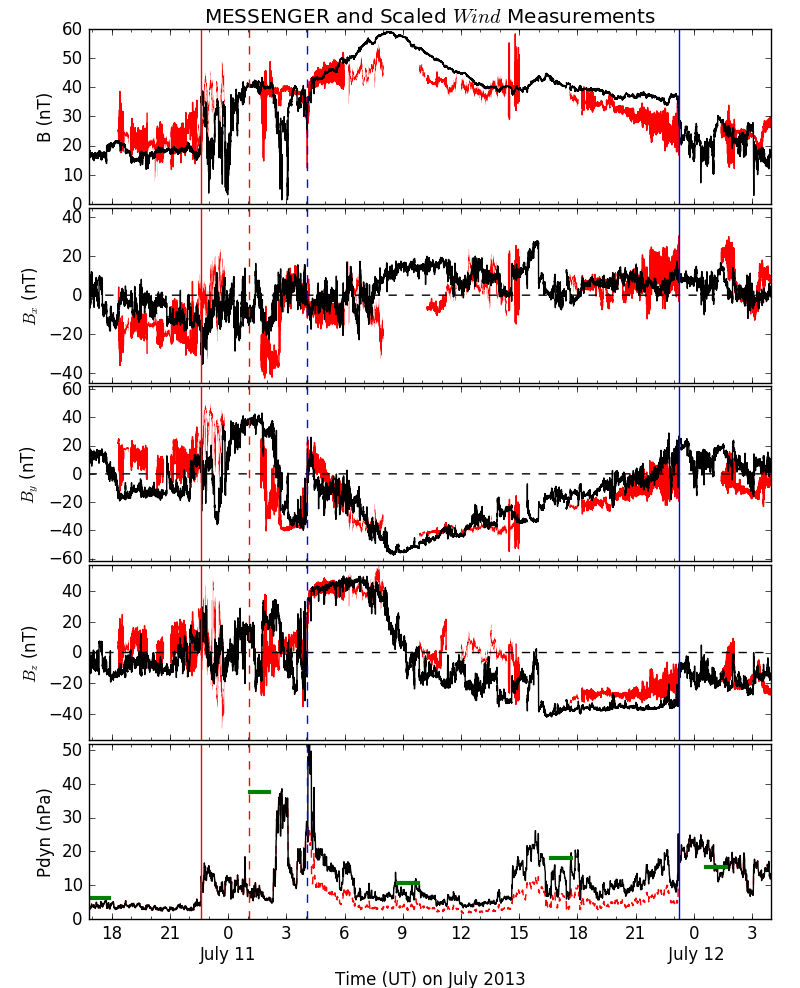}}
\caption{Magnetic field measurements at MESSENGER (red) overlaid with {\it Wind} magnetic field measurements scaled, time-shifted and compressed (black). The last panel shows the dynamic pressure measured at 1 AU scaled as explained in the text and with the same time scaling and shifting as the magnetic field. Green horizontal bars in the last panel indicate the estimate of the dynamic pressure using the outbound Mercury's magnetopause crossings by MESSENGER. The shock arrival time at Mercury is marked with the dashed vertical red line and the scaled and shifted shock arrival from the {\it Wind} data is marked with the solid red line. The dash blue line indicates the end of the sheath and start of the ME, and the solid blue line indicates the end of the ME.}
\label{fig:scaled_plot}
\end{figure}

In Figure~\ref{fig:scaled_plot}, we overlay {\it Wind} and MESSENGER IMF measurements and dynamic pressure estimates. The {\it Wind} magnetic field measurements are scaled using the value of $\alpha_{\max} = -1.6$ found based on the decay of the peak magnetic field between Mercury and Earth (as discussed in the previous section). The same scaling is used for each component. {\it Wind} timing is scaled by a factor of 2.25, corresponding to the duration difference of the ejecta between Mercury and Earth. It is also time-shifted to match the beginning of the ejecta. Theoretical work indicate that the axial field is expected to decrease with distance as $r^{-2}$, whereas poloidal field is expected to decrease as $r^{-1}$ \cite{Farrugia:1993,Leitner:2007,Manchester:2017,Dumbovic:2018}. Here, the same scaling for all components of the magnetic field as presented in Figure~\ref{fig:scaled_plot} gives a much closer match between MESSENGER and {\it Wind} data than a different scaling for the various components. Future work will focus on the change of the magnetic field components with distance for all CMEs measured in conjunction between various spacecraft.

The dynamic pressure measured by {\it Wind} is scaled to MESSENGER's distance as follows: in the solar wind and in the sheath region of the CME, the dynamic pressure is scaled by $r^{-2}$, corresponding to the assumption that mass is conserved as the solar wind expands. In the ejecta, this assumption does not necessarily hold \add{true} due \add{to} the fact that CMEs do not expand self-similarly. Instead, we estimate the dynamic pressure at Mercury inside the ME by using the proton density and speed measured by {\it Wind} and scaled to Mercury's distance. We scale the density measured by {\it Wind} by $r^{-2.44}$ following the finding of \citeA{Leitner:2007} for CMEs measured by Helios. We use the solar wind speed measured at {\it Wind} that we shift up by 50~km\,s$^{-1}$, corresponding to the expected deceleration between Mercury and Earth as determined in Section~\ref{sec:remote}. In Figure~\ref{fig:scaled_plot}, the dynamic pressure scaled as explained here is shown with the solid black line, whereas, inside the ME, the dynamic pressure scaled by $r^{-2}$ is shown with the red dashed line. The dynamic pressure estimates obtained at Mercury from the Newtonian approximation are shown with green horizontal lines. They are obtained by assuming that Mercury's magnetopause is in total pressure balance, so that the IMF magnetic pressure plus a set fraction (here the canonical value of 88\%) of the solar wind dynamic pressure are equal to the magnetic pressure of the magnetosphere at the magnetopause plus the thermal contribution associated with the cusp regions. Further details can be found in Section 2.2 of\citeA{Winslow:2017}. The estimates of the solar wind dynamic pressure from the Newtonian approximation compare favorably with the scaled dynamic pressure measurements, both before the CME hit (around 17 UT on July 10), during the ME (around 9 UT and 17 UT on July 11) and after the end of the CME (around 1 UT on July 12). During the sheath passage (estimate at 1UT on July 11), the Newtonian approximation returns a value of 38 nPa comparable to the peak dynamic pressure estimated around 3 UT. This is most likely associated with the breakdown of the temporal scaling, since the scaling was made for the ME period. This is further discussed below.

Figure~\ref{fig:scaled_plot} highlights that the ME expansion is nearly self-similar between Mercury and Earth i.e., the scaled size of the ME at Earth is nearly identical to the size of the ME at Mercury. Hereafter, we discuss the exceptions to self-similar expansion. First, the total magnetic field (top panel of Figure~\ref{fig:scaled_plot}) profile at Earth appears slightly stronger with potentially a higher peak and slower decay. This may be associated with the expansion at Earth as the time scaling does not take into account that the front of the ejecta is faster than the back by about 20-25$\%$. 

 %%%%%%%%%%%%%%%%%%%%%%%
%%%%%%%%%%%%%%%%%%%%%%%%

 %%%%%%%%%%%%%%%%%%%%%%%
\begin{figure*}[tb]
\centering
{\includegraphics*[width=.98\linewidth]{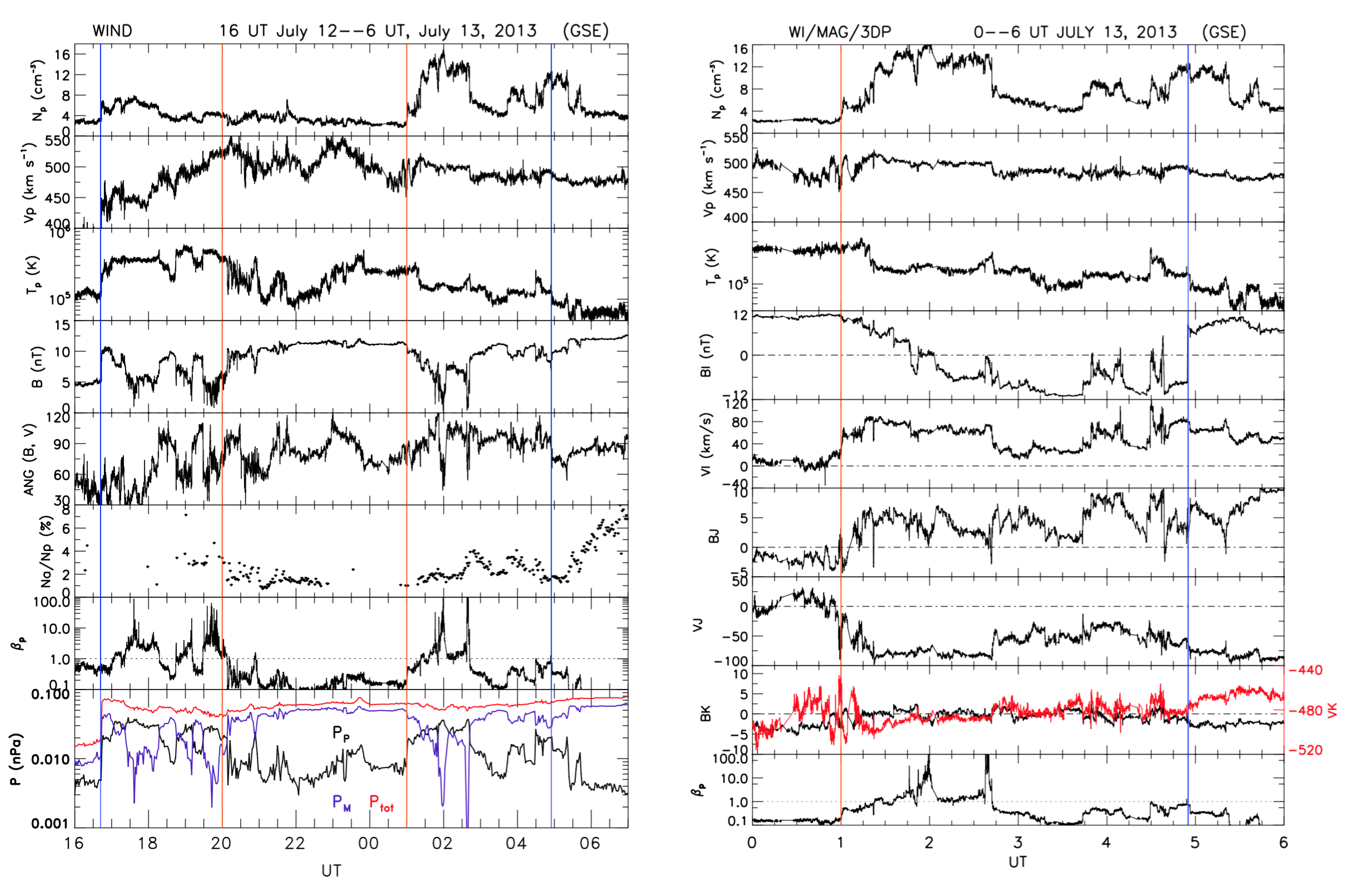}}
\caption{Left: Sheath measurements by {\it Wind}. The panels show, from top to bottom, the proton density (a), velocity (b), temperature (c), the total magnetic field (d), the angle between the magnetic field and flow vectors (e), the $\alpha$-to-proton number density ratio (f), the proton $\beta$ (g) and the pressures (red: total, blue: magnetic, black: proton). Right: MVA analysis of the second half of the sheath showing, from top to bottom, the proton density (a), velocity (b) and temperature (c) and the magnetic field and velocity vectors in MVA $i,j,k$ coordinates (panels d-h with $V_k$ overlaid in red on top of $B_k$ in panel h), as well as the proton $\beta$ (i). The blue vertical lines correspond to the shock arrival time (left) and the end of the sheath (right). The red vertical lines are used to divide the sheath into three parts to guide the discussion in the text.}
\label{fig:MVA}
\end{figure*}
%%%%%%%%%%%%%%%%%%%%%%%%I

{\add Second,} the main deviation from self-similar expansion between Mercury and Earth occurs in the sheath. The scaled {\it Wind} and the MESSENGER measurements of the sheath are shown between the red (dashed or solid) and blue dashed lines in Figure~\ref{fig:scaled_plot}. The shock arrival time at Mercury is marked with the dashed vertical red line whereas the scaled and shifted shock arrival from the {\it Wind} data is marked with the solid red line. If the sheath expansion was self-similar with the same rate as that of the ME expansion, the lines marking the shock arrival at Mercury and Earth would overlap\change{. They do not}{, which is not the case here}. Hereafter, we show that this difference is mainly due to new material being added to the sheath, in addition to the radial expansion of the sheath as a secondary effect. The sheath lasts 2.5 hours at Mercury and 12.2 hours at Earth. The fact that radial expansion occurs in the sheath is made clear by inspecting the planar structure just preceding the start of the ejecta and it appears consistent with the scaled {\it Wind} measurements, i.\,e.\, the expansion rate of this planar structure is similar to that measured for the ME itself. We further discuss the planar structures at the end of this section. 

We now turn our attention to the growth in the sheath between Mercury and Earth. To do so, we first assume that the sheath has the same speed at Mercury as the CME front. With a speed of 500~km\,s$^{-1}$ at Earth and 580~km\,s$^{-1}$ at Mercury, the sheath size is 0.148~AU at Earth {\it vs.} 0.035~AU at Mercury. A radial expansion of the sheath similar to that of the ejecta would result in a sheath of 0.06~AU at Earth, which means that $\sim$ 60\% of the sheath measured at Earth is made up of new material accumulated between Mercury and Earth. 

A potential limitation of this analysis is the fact that the shock is not directly measured in the solar wind by MESSENGER, only its consequence on Mercury's magnetosphere (see Section~\ref{sec:Merc}). In fact, an increase in dynamic pressure measured at Earth (see last panel of Figure~\ref{fig:scaled_plot}) about twice larger than at the shock occurs at a tangential discontinuity around 01:00 UT on July 13 (02:30 UT on July 11 in the time-shifted and scaled plot at the bottom panel of Figure~\ref{fig:scaled_plot}). Note that, in this panel, the dynamic pressure measured at Earth is scaled differently inside and outside the ME. A uniform scaling is obtained by following the black curve outside the ME and the red curve inside the ME. The peak in dynamic pressure occurring between 2:20 UT and 3:05 UT is in fact the largest dynamic pressure measured at Earth.
This increase occurs about $\sim$ 1.3 hours after what we identified as the shock at Mercury. Because we do not have IMF measurements at Mercury during this time period, we cannot fully rule out that what is identified as a shock at Mercury is in fact a sharp increase in dynamic pressure. However, we explain here why we stand by our identification of the shock at Mercury by performing a Gedankenexperiment. Let us assume that what we identified as the CME-driven shock at Mercury corresponds in fact to this dynamic pressure increase in the middle of the sheath, as measured at Earth. At the end of the previous section, we determined that the ME was very likely to drive a shock at Mercury's distance based on Mach number considerations. Therefore, MESSENGER should observe a shock before this dynamic pressure increase, and it should be clearly visible in the measurements when MESSENGER was inside the magnetosheath. MESSENGER was inside the magnetosheath from 22:25UT to 23:57 UT on July 10 (see left panels of Figure~\ref{fig:shock_arrival}), but we see no evidence of such a shock in the measurements. We therefore conclude that the dashed red line in Figure~\ref{fig:scaled_plot} corresponds to the shock arrival time at Mercury.

We now turn to the planar structures in the sheath. We perform a minimum variance analysis on the magnetic field measured at Earth by {\it Wind} from 01 UT to 04:55 UT on July 13. We find that the ratio of middle to lowest eigenvalues is 4.8, indicating a robust result. The magnetic field component normal to the structure is $-0.19 \pm$ 1.05~nT, which is consistent with 0~nT. These two results indicate that this structure is a planar magnetic structure. The normal to the planar structure makes an angle of about $85^\circ$ with the axis of the ME {\it i.e.} the planar structure is almost parallel to the front surface of the ME. This is consistent with the draping of this structure around the ME \cite{Farrugia:2008}. A plot of the measurements at the end of the sheath and beginning of the ejecta in the magnetic ejecta's MVA $i,j,k$ coordinates is shown in the right panel of Figure~\ref{fig:MVA}. %The penultimate panel shows the flows tangent to the ejecta's boundary. These flows increase to above 175~km/s in the last 4 hours of the sheath. This is significantly higher than the expansion speed and significantly larger than Alfven speed in the sheath. 

\subsection{``Parts'' of the Dense Sheath Measured at 1 AU}
The left panel of Figure~\ref{fig:MVA} shows a zoom-in on the sheath at 1~AU with the proton density, velocity, temperature, the total magnetic field, the angle between the magnetic field and flow vectors, the $\alpha$-to-proton number density ratio, the proton $\beta$ and the pressures. Overall, we propose that the first 3.5 hours of the sheath measured at 1~AU (between the blue line that marks the shock and the first red line in the left panel of Figure~\ref{fig:MVA}) is composed of shocked solar wind, as indicated by the high proton temperature and complex magnetic field. This might be associated with a weakening shock as the speed profile is increasing away from the shock.  The next five hours (between the two red lines) are composed of compressed solar wind that accumulated due to the CME propagation and expansion into the solar wind. During this time, the magnetic field (panel d in Figure~7, left) is smooth and elevated but the angle between the magnetic field and the flow (panel e in Figure~7, left) is complex. 

The sheath takes approximately 40 hours to propagate from MESSENGER to {\it Wind}, indicating an average speed of 550~km\,s$^{-1}$. Assuming the solar wind in front of the sheath has a constant speed of 380~km\,s$^{-1}$ (the average value measured at 1~AU over the six hours before the shock arrival), it would take about 60 hours to propagate from MESSENGER to {\it Wind}. There is a difference of 20 hours in the propagation time of the sheath versus the solar wind from Mercury to Earth. This means that the 20 hours of solar wind before the CME-driven shock as measured by MESSENGER are part of the sheath as measured at {\it Wind}. For the three orbits before the CME arrival at MESSENGER, the magnetic field measurements at MESSENGER are predominantly in the $B_y$ positive direction with $B_z$ positive closest to the CME arrival and $B_z$ fluctuating and close to 0 earlier on. These measurements of the solar wind by MESSENGER appear consistent with the {\it Wind} measurements in the sheath, especially the long period of strong and steady $B_y$ positive with more complex and weaker $B_x$ and $B_z$ components.

The last four hours of the sheath (between the last vertical red line and the final vertical blue line in the left panels of Figure~7) is a planar magnetic structure starting with a region of enhanced density (first panel) and dynamic pressure which is already measured at Mercury. This accumulated mass expands with the same profile as the CME, indicating that the CME expansion is the primary driver of the sheath accumulation. Further discussion about the relation between CME expansion and sheath formation can be found in \citeA{Siscoe:2008}. The angle between the magnetic field and the flow is almost 90$^\circ$, further confirming that this material is accumulated. The right panel of Figure~\ref{fig:MVA} shows a zoom-in plot on this last part of the sheath plotted in the MVA coordinates (with the MVA performed between the last red and the last blue vertical lines).

\section{Conclusions}\label{sec:conc}

The 2013 July 9 CME is an event of moderate speed with a leading edge speed at MESSENGER of 580 $\pm$ 30~km\,s$^{-1}$ and at Earth of 500 $\pm$ 30~km\,s$^{-1}$. Remote heliospheric observations confirm that this CME experienced little if any deceleration in the heliosphere. The unusual speed profile in the sheath at 1~AU may be associated with this unusual behavior, as the average speed of the ME at Earth is only slightly faster than the speed of the background solar wind. Even with this detailed analysis, we cannot determine why the second half of the sheath has significantly higher speeds than the front of the ejecta. Such complex behaviors in the sheath are somewhat unusual but not rare.

The CME contains a low-inclined North-West-South \cite{Bothmer:1998,Mulligan:1999} ME, meaning that the field has a southward component during the second half of the ejecta. The magnetic ejecta is large (and of long-duration) at Mercury and Earth. The sheath at Mercury is relatively short, whereas at Earth, it is long and composed of at least two periods. The first period is a typical turbulent sheath behind a shock, and was probably not present at Mercury. Thereafter, there is draped field and a planar magnetic structure with large flows tangent to the magnetic ejecta. This structure is already present at Mercury. It is possible that it formed in the corona or innermost heliosphere, in a process similar to that discussed in \citeA{deForest:2013}. This highlights that the sheath plasma and magnetic field measured between the shock and ME at Earth can be composed of material from various origins: coronal and heliospheric, shocked and compressed, etc.

By investigating the expansion as measured by the decrease in magnetic field between Mercury and Earth and the increase in the ejecta's duration, we determine that this CME was likely large {\it by nature}, meaning that the very long duration at Earth is not the result of extreme heliospheric expansion but is already clearly visible at 0.45~AU. We cannot determine if this large size is a direct result of the initiation mechanism or if the CME expanded strongly in the corona and before reaching Mercury. For this event, we find that the expansion parameter as calculated from {\it in situ} measurements near 1 AU using the formalism of \citeA{Gulisano:2010} is consistent with the magnetic field exponent decrease between Mercury and Earth. It is also consistent with the findings from a recent study by \citeA{Al-Haddad:2019} using numerical simulations to investigate the expansion of CMEs and associated fall-off of their magnetic field strength with distance in the innermost heliosphere. If the relation between expansion parameter and magnetic field exponent decrease is confirmed, this further justifies the paradigm of \citeA{Gulisano:2010}, which theoretically relates these two quantities. This would make it possible to determine the local CME expansion from plasma measurements close to the Sun, by Parker Solar Probe, Solar Orbiter or future Sentinel missions and to use this as a proxy for the global expansion of the ME. It would then be possible to forecast how the magnetic field inside the ME at Earth from the measurements in the innermost heliosphere. This case study however needs to be followed up by an investigation with a more statistical approach to determine if this result holds true in general.

From the combination of MESSENGER data, including estimates of the dynamic pressure from magnetopause crossing, and scaled {\it Wind} data, we also determine the magnetic field and dynamic pressure profile inside the CME at Mercury. We confirm the typical structure of a CME also at Mercury: a sheath with a high dynamic pressure followed by an ejecta with low dynamic pressure, in this case with lower dynamic pressures in the first half and higher dynamic pressures in the second half. The end of the CME is marked by an increase in dynamic pressure. Overall, this indicates that, except for its long-duration at Mercury and at Earth, this is a medium-speed CME with overall typical characteristics: the average magnetic field inside the ejecta at 1~AU is almost identical to the average (12.6 nT) for all magnetic clouds as reported by \citeA{Richardson:2010}, the maximum ME magnetic field and CME speed measured at Earth is comparable to that of all CMEs measured by STEREO during solar cycle 24 \cite{Jian:2018}, the expansion parameter is typical and the expansion speed is in-between the average values reported by \citeA{Jian:2018} (62~km\,s$^{-1}$) and \citeA{Richardson:2010} (43~km\,s$^{-1}$). In addition, the duration of the sheath as compared to that of the ejecta at Mercury and Earth are consistent with past studies \cite{Jian:2018,Janvier:2019}. Overall, this IMF data and solar wind estimates can then be used to analyze the magnetospheric measurements by MESSENGER in order to compare the effect\add{s} of the CME on Mercury and Earth's magnetosphere. We plan to perform such an investigation in a follow-up study.

\acknowledgments

The authors acknowledge the use of MESSENGER data available at Planetary Data System (\url{https://pds.nasa.gov/}) and NASA/GSFC's Space Physics Data Facility's CDAWeb service for {\it Wind} data available at CDAWeb (\url{https:cdaweb.gsfc.nasa.gov/index.html/}). STEREO/HI measurements were obtained from the STEREO RAL website (\url{https://www.stereo.rl.ac.uk}). Results from the database cited in the manuscript come from DONKI (\url{https://swc.gsfc.nasa.gov/main/donki}), HELCATS (\url{https://www.helcats-fp7.eu/catalogues/wp3_cat.html}), SEEDS (\url{http://spaceweather.gmu.edu/seeds/secchi.php}), and the CDAW LASCO catalog (\url{https://cdaw.gsfc.nasa.gov/CME_list}).

N.~L acknowledges support from NASA grants NNX15AB87G, 80NSSC19K0831 and NSF grant AGS1435785. R.~M.~W. acknowledges support from NSF grant AGS1622352, and NASA grants NNX15AW31G and 80NSSC19K0914. C.~J.~F. acknowledges support from NNX16AO04G and 80NSSC19K1293.

%% ------------------------------------------------------------------------ %%
%% References and Citations

%%%%%%%%%%%%%%%%%%%%%%%%%%%%%%%%%%%%%%%%%%%%%%%
%
% \bibliography{<name of your .bib file>} don't specify the file extension
%
% don't specify bibliographystyle
%%%%%%%%%%%%%%%%%%%%%%%%%%%%%%%%%%%%%%%%%%%%%%%

%Reference citation instructions and examples:
%
% Please use ONLY \cite and \citeA for reference citations.
% \cite for parenthetical references
% ...as shown in recent studies (Simpson et al., 2019)
% \citeA for in-text citations
% ...Simpson et al. (2019) have shown...
%
%
%...as shown by \citeA{jskilby}.
%...as shown by \citeA{lewin76}, \citeA{carson86}, \citeA{bartoldy02}, and \citeA{rinaldi03}.
%...has been shown \cite{jskilbye}.
%...has been shown \cite{lewin76,carson86,bartoldy02,rinaldi03}.
%... \cite <i.e.>[]{lewin76,carson86,bartoldy02,rinaldi03}.
%...has been shown by \cite <e.g.,>[and others]{lewin76}.
%
% apacite uses < > for prenotes and [ ] for postnotes
% DO NOT use other cite commands (e.g., \citeA, \citep, \citeyear, \nocite, \citealp, etc.).
%

\end{document}